# Optical elements containing semitrasparent wavelike films

ANATOLY M. SMOLOVICH[1,*] AND VALERY CHERNOV[2]

[1]*Kotel'nikov Institute of Radio Engineering and Electronics (IRE) of the Russian Academy of Sciences, Mokhovaya 11-7, Moscow 125009, Russia*
[2]*Departamento de Investigación en Física, Universidad de Sonora, AP 5-088, Hermosillo, Sonora 83190, Mexico*
*\*Corresponding author: asmolovich@petersmol.ru*

**Novel optical elements containing semitransparent wavelike films embedded into the bulk of transparent material, which form a reflection image without transmitted light distortion, are studied. The dynamic theory of light diffraction by a locally periodic multilayer semitransparent wavelike film is developed. A simple analytical formula for near Bragg diffraction order intensity is obtained for the case when only one diffraction order lies within the hologram angular selectivity. The phase modulation of light transmitted through the optical element containing wavelike films is estimated for single-layer and multilayer wavelike films with an arbitrary shape of surface. The restrictions on the structure parameters for which transmitted light distortions would be negligible are obtained. A new type of high quality color hologram is proposed and shown to be feasible by calculation of hologram diffraction efficiency and spectral selectivity for three colors. Other possible applications, such as monochrome and color head-up and head-mounted displays, and imaging on spectacle lenses, are discussed.**

## 1. INTRODUCTION

In some cases, it is necessary to reduce distortion of light transmitted through a hologram. For example, this is important if a reflection hologram is used as a screen (also called combiner) in a head-up display of a vehicle [1, 2] when possibility of road observing through the screen is necessary. Several different types of head-up displays have been known for many years and are currently used mostly in aviation. The development of a simple and inexpensive head-up display for automobile applications, however, is still of interest. Recently the mainstream use of car navigation systems has created a crash risk as a driver has to frequently turn his/her eyes from the road to the navigation device screen and also to change his/her eyes accommodation from infinity onto a short distance. To avoid this risk head-up displays, which form an image in infinity over the road view, should be used.

In a similar way a hologram is used in a head-mounted display (also called a head-worn display or wearable glasses) to display a see-through image imposed upon a real world view, creating what is called augmented reality [3, 4]. The real world view is seen through the surface of the display. Currently this technique of information displaying became a very promising direction of development as an alternative to the mobile device displays such as the ones of cellular phones, pocket computers, TV and video players due to their limitation in ability to increase the volume and quality of visual information and to be compact at the same time. The use of head-mounted displays could solve this problem.

In conventional reflection holograms with surface phase relief, different beams passing through a hologram have different optical path lengths (OPLs). This results in a phase modulation of light passing through the hologram. Hence, distortion of the transmitted wave takes place. In volume holograms, the distortion of transmitted light is notably lower. However, volume holograms could not be manufactured by low-cost embossing technology [5, 6]. As a result, the production of these holograms is expensive.

Transmitted light distortion can be reduced by the use of a special type of optical element (OE) described below. This OE (Fig. 1 (a)) contains a semitransparent wavelike film *1* embedded into transparent material *2*. This wavelike film has a uniform thickness and is confined between identical relief surfaces. The semitransparent wavelike film reflects a part of light incident to the OE, while some light passes through the OE. The OPLs of the light transmitted through the OE in different areas are practically equal because the refractive index of the transparent material on both sides of the wavelike film is the same. We will call the holographic (locally periodic) semitransparent wavelike film relief shape as the wavelike hologram. Thus, the wavelike hologram forms an image in reflected light similarly to the ordinary hologram with surface relief. But in contrary to the former the part of reconstructing beam transmitted through the wavelike hologram does not have significant phase distortion. A semitransparent wavelike film inside the OE can also have the chaotic diffuse relief shape. The diffuse wavelike film also should have the uniform thickness and should be confined between identical relief surfaces (Fig. 1 (b)). Another type of wavelike film consists of several parallel wavelike layers made of different materials with different indexes of refraction (Fig. 1 (c)). This is a multilayer film. The multilayer film will have a spectral selectivity. This allows its use in color devices, for example, color displays.

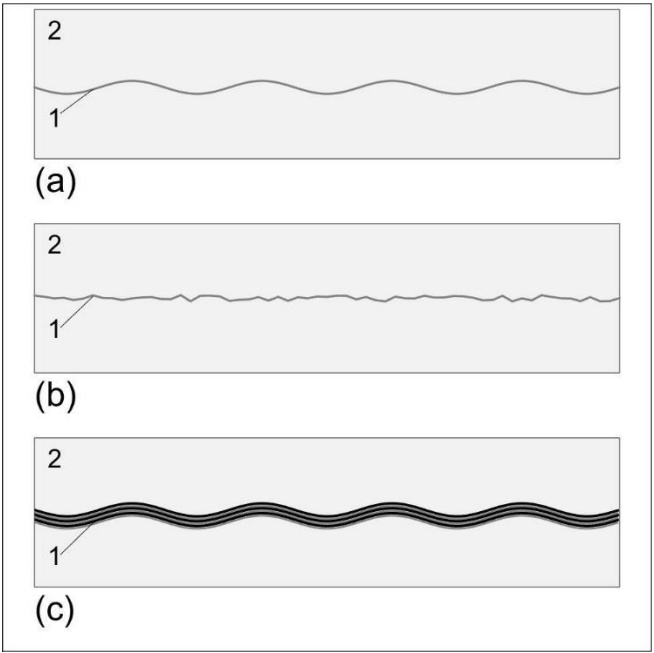

Fig. 1. Different types of semitransparent wavelike films (*1*) embedded into the bulk of OE (*2*): (a) Single layer holographic wavelike film. (b) Single layer chaotic diffuse relief film. (c) Multilayer holographic wavelike film.

The basic idea of the OE containing semitransparent wavelike films, which do not distort transmitted light, has been used earlier in several applications [7-9]. Among them are the read-only optical disks with several semitransparent information layers containing tracking grooves and information pits [10, 11]. When information is being read from one of the optical disk layers the other layers do not distort the reading and signal beams. An exotic example of the OE containing semitransparent thin film which does not distort transmitted light is an interference structure recorded in volume medium by counterpropagating femtosecond laser pulses. Such structures demonstrate the geometrical-optical mechanism of wavefront reconstruction different from the holographic one [12, 13].

However, to the best of our knowledge, the distortion of light transmitted through the OE containing holographic and chaotic diffuse relief films (Fig. 1 (a), (b), and (c)) was not specially investigated. It is obvious that some restrictions on the wavelike film parameters including its thickness must be applied to provide acceptably low level of phase modulation of transmitted light. It is important to understand what values of the wavelike hologram diffraction efficiency or/and spectral selectivity could be achieved under these restrictions. In Section 3, we will show that the restrictions mentioned above do not impose any difficulties on a single-layer wavelike film. This is not the case, however, for a multilayer wavelike film because its thickness is substantially higher. This situation should be investigated. To the best of our knowledge there is no theory specially developed for multilayer wavelike holograms (Fig. 1 (c)). We believe, however, that the model proposed in [14] for X-ray-diffraction by a multilayer structure modulated by surface acoustic-waves may be applied here. Unfortunately, this model is a kinematic one. The kinematic approximation ignores attenuation of the incident wave and secondary scattering of the waves diffracted by each layer. It can give the correct value of the diffracted orders amplitudes only if they are significantly lower than the incident wave amplitude. This approximation cannot be used for the most interesting case of practical use when the diffraction efficiency of the structure is high. We believe that the methods developed for multilayer-coated diffraction gratings [15-18] can also be used here. Unfortunately, as far as we know, these methods do not give a simple analytical formula for diffraction efficiency, which can be used in practice. Such a formula is strongly desirable for the transmitted light distortion estimation. Note, that later the universal approach to reflection by the volume Bragg structures was developed in [19, 20].

Thus, our purpose is to obtain an analytical expression for the diffraction efficiency of the semitransparent multilayer wavelike film in the dynamic approximation and to study its parameters restrictions necessary for acceptably low level of phase modulation of transmitted light. In Section 2, we develop the dynamic theory of the multilayer wavelike hologram (Fig. 1 (c)). In Section 3, we estimate the phase modulation of light transmitted through the OE containing wavelike films. Firstly, we consider a single layer film with an arbitrary shape of surface (including the ones shown in Fig. 1 (a) and (b)). Then we generalize this estimation for the case of the multilayer wavelike film (Fig. 1 (c)). In Section 4, we discuss the possible applications of the OEs containing wavelike films and the manufacturing methods for their production. We propose a new type of color hologram. The results obtained in Sections 2 and 3 are used to investigate its principal possibility of realization.

## 2. DIFFRACTED REFLECTED BEAMS: DYNAMIC THEORY

Let us consider multilayer wavelike structure containing some number of alternate layers of two different materials with dielectric constants $\varepsilon^{(1)}$ and $\varepsilon^{(2)}$, conductivities $\sigma^{(1)}$ and $\sigma^{(2)}$, and thicknesses $d_1$ and $d_2$, respectively. The structure is oriented normally to the *z*-axis (Fig. 2). The structure period in *z*-direction is $\Delta = d_1 + d_2$. The surface of each layer is periodically modulated along the *x*-axis by the law:

$$\Psi(x) = h \sin\left(\frac{2\pi x}{\Lambda}\right), \tag{1}$$

where $h$ and $\Lambda$ are, respectively, amplitude and period of the periodic along axis *x* wavelike disturbance of the multilayer structure. Let us write the local dielectric constant of the multilayer structure as a function of *x* and *z*: $\varepsilon = \varepsilon(x, z)$. If $h = 0$, $\varepsilon$ does not depend on *x* and is a step function of *z*: $\varepsilon = \varepsilon(z)$, which has two values $\varepsilon^{(1)}$ and $\varepsilon^{(2)}$. Let us describe the periodic structure modulation along the *x*-axis as the *x*-dependent shift of the $\varepsilon(z)$ function argument:

$$\varepsilon(x,z) = \varepsilon\big(z - \Psi(x)\big). \tag{2}$$

Using the approach [21-22] we expand (2) into Fourier series

$$\varepsilon(z) = \sum_{n=-\infty}^{\infty} b_n \exp(inPz), \tag{3}$$

where

$$b_n = \frac{1}{\Delta} \int_{-\Delta/2}^{\Delta/2} \varepsilon(z)\exp(-inPz)dz.$$

Confining ourselves to zero-order and ±1st orders:

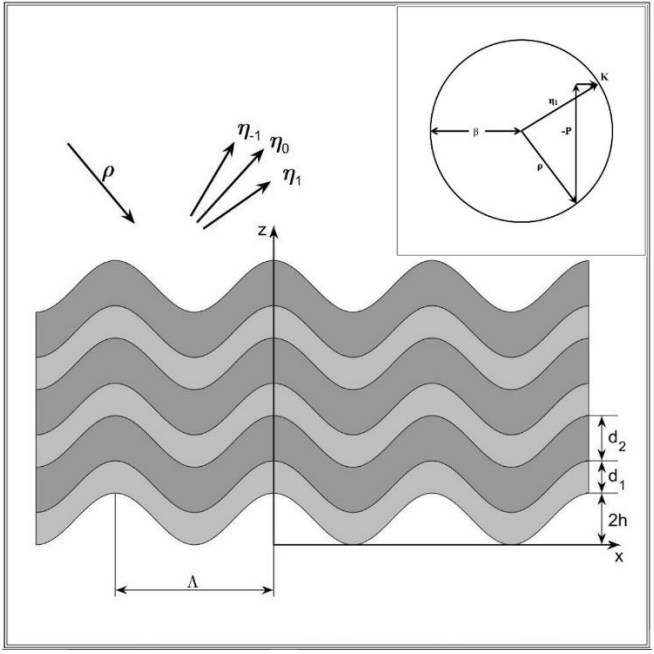

Fig. 2. Multilayer wavelike structure. $d_1$ and $d_2$ are thicknesses of layers, $\Delta$ is the period of multilayer structure along axis $z$: $\Delta = d_1 + d_2$, $h$ and $\Lambda$ are, respectively, the amplitude and period of the periodic along axis $x$ wavelike disturbance of the multilayer structure, $\boldsymbol{\rho}$ and $\boldsymbol{\eta_m}$ are the wave vectors of the incident and $m$-th order diffracted reflected waves. Insert: Vector diagram for the case of the Bragg condition fulfilled for the first diffraction order, $\boldsymbol{P}=P\boldsymbol{e_z}$ is a vector of the grating related to multilayer structure along axis $z$, where $P=2\pi/\Delta$, $\boldsymbol{K}=K\boldsymbol{e_x}$ is a vector of the grating related to periodic along axis $x$ wavelike disturbance of the multilayer structure, where $K=2\pi/\Lambda$, $\beta=2\pi\varepsilon_0^{1/2}/\lambda$, $\varepsilon_0$ is the average dielectric constant, $\lambda$ is the wavelength of light.

$$\varepsilon = \varepsilon_0 + \frac{\varepsilon_1}{2}\exp\{iP[z-\Psi(x)]\} + \frac{\varepsilon_1}{2}\exp\{-iP[z-\Psi(x)]\} = \varepsilon_0 + \frac{\varepsilon_1}{2}\exp(iPz)\exp[-iPh\sin(Kx)] + \frac{\varepsilon_1}{2}\exp(-iPz)\exp[iPh\sin(Kx)], \quad (4)$$

where $P = |\boldsymbol{P}|=2\pi/\Delta$, $K=|\boldsymbol{K}| = 2\pi/\Lambda$, $\boldsymbol{P}$ and $\boldsymbol{K}$ are the vectors of gratings along axes $z$ and $x$ respectively, $\varepsilon_0$ is the average dielectric constant (zero-order coefficient of Fourier series):

$$\varepsilon_0 = \frac{\varepsilon^{(1)}d_1 + \varepsilon^{(2)}d_2}{d_1+d_2}, \quad (5)$$

$\varepsilon_1$ is the first-order coefficient of Fourier series:

$$\varepsilon_1 = \frac{2(\varepsilon^{(1)}-\varepsilon^{(2)})}{\pi}\sin\left(\frac{\pi d_1}{d_1+d_2}\right). \quad (6)$$

Let us use the well-known expansion [23]:

$$\exp[ia\sin(b)] = \sum_{m=-\infty}^{\infty} J_m(a)\exp(imb), \quad (7)$$

where $J_m(a)$ is the $m$-th order Bessel function. We have:

$$\varepsilon = \varepsilon_0 + \frac{\varepsilon_1}{2}\exp(iPz)\sum_{m=-\infty}^{\infty} i^m J_m(-Ph)\exp(imKx) + \frac{\varepsilon_1}{2}\exp(-iPz)\sum_{m=-\infty}^{\infty} i^m J_m(Ph)\exp(imKx) \quad (8)$$

Let us use the Bessel function property:

$$J_m(-a) = (-1)^m J_m(a) \quad (9)$$

$$\varepsilon = \varepsilon_0 + \frac{\varepsilon_1}{2}\exp(iPz)\sum_{m=-\infty}^{\infty}(-1)^m J_m(Ph)\exp(imKx) + \frac{\varepsilon_1}{2}\exp(-iPz)\sum_{m=-\infty}^{\infty} J_m(Ph)\exp(imKx) \quad (10)$$

We can also expand the structure local conductivity $\sigma$ in similar way:

$$\sigma = \sigma_0 + \frac{\sigma_1}{2}\exp(iPz)\sum_{m=-\infty}^{\infty}(-1)^m J_m(Ph)\exp(imKx) + \frac{\sigma_1}{2}\exp(-iPz)\sum_{m=-\infty}^{\infty} J_m(Ph)\exp(imKx), \quad (11)$$

where $\sigma_0$ and $\sigma_1$ are the average and the first-order Fourier coefficient of the structure local conductivity respectively.

Let us consider a linearly polarized plane wave with the wave vector $\boldsymbol{\rho}$ lying in the $xz$-plane and the electric field vector $\boldsymbol{E}$ normal to the $xz$-plane falling onto the structure. Wave propagation in the structure is described by the scalar wave equation [24, 25]:

$$\nabla^2 E + k^2 = 0, \quad (12)$$

where $E(x, z)$ is the complex amplitude of the $y$-component of the electric field, which is assumed to be independent from $y$ and to oscillate with an angular frequency $\omega$. The propagation constant $k(x, z)$ is spatially modulated due to modulation of the dielectric constant $\varepsilon$ and the conductivity $\sigma$ of the structure:

$$k^2 = \frac{\omega^2}{c^2}\varepsilon - i\omega\mu\sigma, \quad (13)$$

where $c$ is the light velocity in free space and $\mu$ is the permeability of the structure. Further we assume that $\mu=1$. After substitution of (10-11) to (13):

$$k^2 = \beta^2 - 2i\alpha\beta + 2\chi\beta[\exp(iPz) \times \sum_{m=-\infty}^{\infty}(-1)^m J_m(Ph)\exp(imKx) + \exp(-iPz)\sum_{m=-\infty}^{\infty} J_m(Ph)\exp(imKx)], \quad (14)$$

where $\beta = \varepsilon_0^{1/2}\left(\frac{\omega}{c}\right) = \frac{2\pi\varepsilon_0^{1/2}}{\lambda}$, $\alpha$ is the average adsorption coefficient of the structure: $\alpha = \frac{\mu c \sigma_0}{\varepsilon_0^{1/2}}$.

The coupling constant $\chi$ is defined as:

$$\chi = \frac{1}{4}\left[\frac{2\pi\varepsilon_1}{\lambda\varepsilon_0^{1/2}} - \frac{i\mu c\sigma_1}{\varepsilon_0^{1/2}}\right]. \quad (15)$$

Similar to [25] let us search solution of equation (12-15) in a form:

$$E = R(z)\exp(-i\boldsymbol{\rho}\boldsymbol{r}) + \sum_{m=-\infty}^{\infty} S_m(z)\exp(-i\boldsymbol{\eta_m}\boldsymbol{r}), \quad (16)$$

where $\boldsymbol{r}$ is the radius-vector, $\boldsymbol{\rho}$ and $\boldsymbol{\eta_m}$ are the wave vectors of the incident and diffracted waves, respectively (Fig. 2),

$$\eta_m = \rho + Pe_z + mKe_x, \qquad (17)$$

$e_x$ and $e_z$ are the unit vectors along the axes $x$ and $z$, respectively. We assume that vector $P$ is directed oppositely to $e_z$ (so, $P_z=-P$) and vector $K$ is directed along $e_x$. Also, following [25] we consider (-1) diffraction order by periodic structure along axis $z$. The slowly varying amplitudes $R(z)$, $S_0(z)$, and $S_m(z)$ are related to the incident, mirror reflected (more exactly, this is the order diffracted only by the periodic structure along the axis $z$) and $m$-th order diffracted waves, respectively. Let us consider the case when

$$\delta\theta < |\theta_n - \theta_{n-1}|, \qquad (18)$$

where $\delta\theta$ is the structure angular selectivity, $\theta_n$ is the incident beam angle with axis $z$, for which the Bragg condition is fulfilled for $n$-th diffraction order. This allows taking into account only an incident wave and one diffraction order most close to the Bragg condition. The incident angle can be chosen to approximately fulfill the Bragg condition for one of the diffracted waves $S_n$: $\eta_n^2 \approx \beta^2$. This is illustrated in Insert of Fig. 2 by the vector diagram for case of $n=1$. In [14] it was shown that in this case in the kinematic approximation the amplitudes of all waves in (16) except $R$ and $S_n$ were small. So, we neglect these waves. Following approach [25] we set the sums of factors of exponents $exp(-i\rho r)$ and $exp(-i\eta_n r)$ equal to zero and neglect the terms containing other exponents as they are related to the waves, which are far from the Bragg condition. In addition, we neglect the components containing the second derivative of the slowly varying amplitudes. In result, we obtain the system of two coupled wave equations:

$$c_R \frac{\partial R}{\partial z} + \alpha R = -i\chi J_n(Ph) S_n$$
$$c_n \frac{\partial S_n}{\partial z} + (\alpha + i\Gamma_n) S_n = -i\chi J_n(Ph) R, \qquad (19)$$

where $c_R = \frac{\rho_z}{\beta}$, $c_n = \frac{\eta_{nz}}{\beta}$ $(c_n \leq 0)$, $\Gamma_n = \frac{\beta^2 - \eta_n^2}{2\beta}$.

$\Gamma_n$ defines the deviation from the Bragg condition. The boundary conditions are similar to the reflection hologram boundary conditions in [25]: $R(0) = 1$, $S_n(T) = 0$, where $T$ is the structure thickness. We do not take into account the Fresnel reflection on the structure boundary.

After solution of (19) similarly [25] we obtain the amplitude of diffracted order:

$$S_n(0) = -i\chi J_n(Ph)$$
$$\left\{ \alpha + i\Gamma_n + \frac{c_n[\gamma_1 \exp(\gamma_2 T) - \gamma_2 \exp(\gamma_1 T)]}{\exp(\gamma_2 T) - \exp(\gamma_1 T)} \right\}^{-1}, \qquad (20)$$

where

$$\gamma_{1,2} = -\frac{1}{2}\left(\frac{\alpha}{c_R} + \frac{\alpha}{c_n} + i\frac{\Gamma_n}{c_n}\right) \pm$$
$$\frac{1}{2}\left[\left(\frac{\alpha}{c_R} - \frac{\alpha}{c_n} - i\frac{\Gamma_n}{c_n}\right)^2 - \frac{4\chi^2 J_n^2(Ph)}{c_R c_n}\right]^{\frac{1}{2}}. \qquad (21)$$

The optimal value of $h$ corresponds to the first maximum of the Bessel function $J_n(Ph)$ in (20). The diffraction efficiency $DE_n$ in the $n$-th diffraction order is defined similarly to expression (40) in [25]:

$$DE_n = \frac{|c_n| S_n(0) S_n^*(0)}{c_R}, \qquad (22)$$

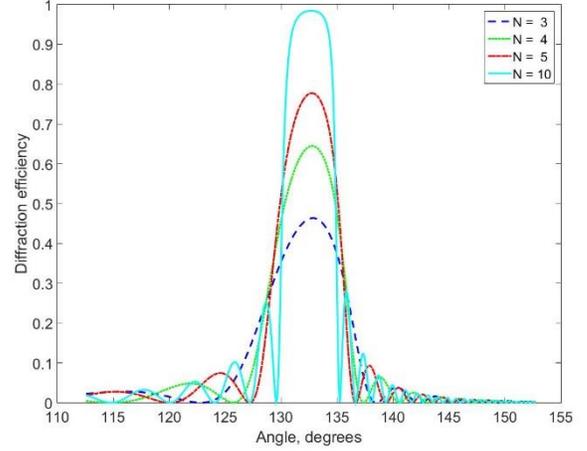

Fig. 3. Calculated diffraction efficiency $DE_1$ of the first diffraction order of the lossless dielectric multilayer structure as a function of the angle between reconstructing wave vector and axis $z$ (angular selectivity) for several specific amounts of $N$. $N$ is the number of the structure periods in z-direction ($N=T/\Lambda$). The parameters of the structure are the following: $\varepsilon^{(1)} = 2.5$, $\varepsilon^{(2)} = 2.0$, $d_1=d_2=2.11\times10^{-7}$ m, $\Lambda=2\times10^{-6}$ m, cosines of the incident and diffracted waves are $c_R = 0.6775$, $c_1 = -0.3225$, $h=1.23\times10^{-7}$ m.

where the asterisk means complex conjugation. In exact Bragg case for $n$-th diffraction order $\Gamma_n = 0$. If additionally the structure is lossless, the expressions (20-22) become simpler:

$$DE_n = \tanh^2\left[\frac{\chi J_n(Ph)}{(c_R |c_n|)^{1/2}} T\right], \qquad (23)$$

where $tanh(x)$ is hyperbolic tangent. We see that in this case the diffraction efficiency in the $n$-th diffraction order approaches the 100% asymptotically with growth of $T$.

We can use the conditions for the theory validity discussed in chapter 6 of [25] because we use the same approach. The main criterion mentioned there is $Q>>1$ where $Q = 2\pi\lambda T/\varepsilon_0^{1/2}\Lambda^2$. $Q$ is called Klein-Cooke parameter. For our case the criterion is satisfied. The additional condition of validity of equations (20-23) is inequality (18) discussed above.

The calculated diffraction efficiency $DE_1$ of the first diffraction order of the lossless dielectric multilayer structure is shown in Fig. 3 for illustration of the theory. $N$ is the number of the structure periods in z-direction ($N=T/\Lambda$). The parameters of the structure are the following: $\varepsilon^{(1)} = 2.5$, $\varepsilon^{(2)} = 2.0$, $d_1=d_2=2.11\times10^{-7}$ m, $\Lambda=2\times10^{-6}$ m, $h=1.23\times10^{-7}$ m, cosines of the incident and diffracted waves are $c_R = 0.6775$, $c_1 = -0.3225$. Diffraction efficiency is shown for several specific amounts of $N$ as a function of the angle between reconstructing wave vector and axis $z$ (angular selectivity). Using these results, it can be demonstrated that condition (18) is fulfilled for $N\geq 3$ here.

The question remains: How does the wavelike structure with these parameters distort transmitted light? The estimation of the phase modulation of light transmitted through the structure is provided in the next section of this paper.

## 3. TRANSMITTED LIGHT: ESTIMATION OF PHASE MODULATION

Fig. 4 shows a magnified fragment of a single-layer wavelike film *1* (shown as a plane-parallel plate) with refractive index $n_1$

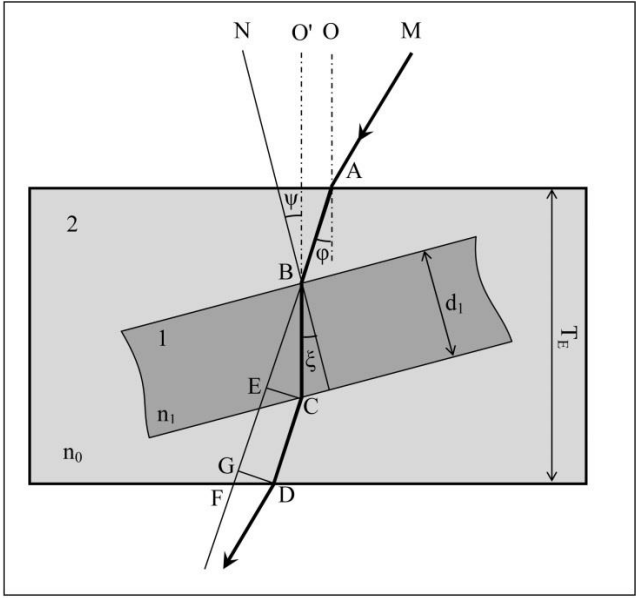

Fig. 4. Magnified fragment of the single layer wavelike film (*1*) with refractive index $n_1$ inside transparent material (*2*) with refractive index $n_0$ of the OE. The incident beam *M* inside OE forms an angle $\varphi$ with the line *O* normal to the surface of the OE. *M* intersects the surfaces of the OE and wavelike film in points *A, B, C,* and *D*. The local normal *N* to the surface of the wavelike film in point *B* forms an angle $\psi$ ($\psi <0$) with the line *O'* normal to the surface of the OE (*O'* is parallel to *O*). $\xi$ is the angle between the light beam inside the wavelike film and its local normal. *F* is the point of intersection of prolongation of the incident beam *M* inside OE with lower surface of the OE. *BE* and *EG* are the projections of *BC* and *CD* onto *AF* respectively. $T_E$ is the thickness of the OE, $d_1$ is the thickness of the wavelike film.

inside the OE made of transparent material *2* with refractive index $n_0$. Within some small area, a ray-tracing approach can be used [26]. The direction of incident beam *M* inside the OE forms an angle $\varphi$ with the line *O* normal to the surface of the OE shown by a dashed line. Here points *A, B, C,* and *D* are the points of intersection of the incident beam (including its refraction) with the surfaces of the OE and wavelike film. The local normal *N* to the surface of the wavelike film at the point *B* forms an angle $\psi$ with the line *O'* normal to the surface of the OE (*O'* is parallel to *O*). We measure the angles $\varphi$ and $\psi$ clockwise taking into consideration their signs. For example, $\psi$ in Fig. 4 is negative. The angle between the direction of the beam and the normal *N* is equal to the difference of these angles $\varphi$ - $\psi$, $\xi$ is the angle between the incident beam inside the wavelike film and its local normal, which can be found from Snell's law:

$$\sin \xi = \frac{n_0}{n_1} \sin(\varphi - \psi). \qquad (24)$$

*F* is the point of intersection of prolongation of the incident beam *M* inside OE with lower surface of the OE. *BE* and *EG* are the projections of *BC* and *CD* onto *AF* respectively. We can consider ray propagation through the wavelike film as propagation through the plane-parallel plate if the ray shift during propagation is much less than the period of the grating (1) along *x*-axis:

$$d_1 \tan \xi \ll \Lambda, \qquad (25)$$

where $d_1$ is the thickness of the wavelike film. In case of non-regular relief $\Lambda$ should be replaced by the typical size of granularity. The OPL *L* of the beam passing through the OE can be calculated from the elementary geometry:

$$L = (AB+CD)n_0 + BC n_1 = \left(\frac{T_E}{\cos\varphi} - BE - GF\right)n_0 + \frac{d_1 n_1}{\cos\xi}, \qquad (26)$$

where $T_E$ is the thickness of the OE,

$$BE = BC\cos(\angle EBC) = BC \times \cos(\varphi-\psi-\xi) = \frac{d_1}{\cos\xi}\cos(\varphi-\psi-\xi), \qquad (27)$$

$$GF = DG\tan(\varphi) = BC\sin(\angle EBC)\tan(\varphi) = BC\sin(\varphi-\psi-\xi)\tan(\varphi) = \frac{d_1}{\cos\xi}\sin(\varphi-\psi-\xi)\tan(\varphi).$$

After substitution (27) into (26):

$$L = \left(\frac{T_E}{\cos\varphi} - \frac{d_1\cos(\varphi-\psi-\xi)}{\cos\xi} - \frac{d_1\sin(\varphi-\psi-\xi)\tan\varphi}{\cos\xi}\right) \times n_0 + \frac{d_1 n_1}{\cos\xi}. \qquad (28)$$

Let us suppose that the angles $\varphi$ and $\psi$ are small and let us use the approximations:

$$\xi = \frac{n_0}{n_1}(\varphi-\psi), \qquad (29)$$

$$\frac{1}{\cos\xi} = (1-\sin^2\xi)^{-\frac{1}{2}} = 1+\frac{\xi^2}{2}, \qquad (30)$$

$$\cos(\phi-\psi-\xi) = [1-\sin^2(\phi-\psi-\xi)]^{\frac{1}{2}} = 1 - \frac{1}{2}(\phi-\psi)^2 + \xi(\phi-\psi) - \frac{\xi^2}{2}. \qquad (31)$$

After substitution (29-31) into (28), replacing *sin* and *tan* in (28) by their arguments, and neglecting the components containing the angles $\varphi$ and $\psi$ in power higher than second we obtain:

$$L = \frac{T_E n_0}{\cos\varphi} + d_1(n_1-n_0) - \frac{d_1 n_0(n_1-n_0)(\varphi^2-\psi^2)}{2n_1}. \qquad (32)$$

The OPL *L* varies for different rays of the beam from $L_{min}$ to $L_{max}$ due to varying of the local surface normal angle $\psi$. The maximum OPL difference $\Delta L$ is equal:

$$\Delta L = L_{max} - L_{min} = \frac{d_1 n_0(n_1-n_0)}{2n_1}[(|\varphi^2-\psi^2|)_{max} - (|\varphi^2-\psi^2|)_{min}]. \qquad (33)$$

The maximum phase modulation $\delta$ due to passing of the plane beam through the OE wavelike film is:

$$\delta = \frac{2\pi}{\lambda}\Delta L. \qquad (34)$$

The level of distortion of transmitted light is determined by the value of $\delta$. Let us consider the level of distortion to be negligible when $\delta$ is less than $b\pi$, where *b* is coefficient determining the required transmitted light quality level. As an example *b* can be equal to 1/20. So, the condition of negligible distortion can be written as:

$$\delta = \frac{\pi d_1 n_0(n_1-n_0)}{\lambda n_1}[(|\varphi^2-\psi^2|)_{max} - (|\varphi^2-\psi^2|)_{min}] < b\pi. \qquad (35)$$

For a normal beam incidence $\varphi = 0$ and $(\psi)^2{}_{min} = 0$. Thus we obtain:

$$\delta = \frac{\pi d_1 n_0(n_1-n_0)}{\lambda n_1}(\psi)^2_{max}. \qquad (36)$$

Let us apply the estimation (36) to the case of sinusoidal shape (1) of the wavelike film. The maximum of $|\psi|$ can be found as the maximum of the module of derivative of (1):

$$|\psi|_{max} = \frac{2\pi h}{\Lambda}. \qquad (37)$$

After substitution (37) into (36):

$$\delta = \frac{4\pi^3 h^2 d_1 n_0 (n_1 - n_0)}{\Lambda^2 \lambda n_1}. \qquad (38)$$

Metal is the most suitable and efficient material for a single-layer wavelike film. To provide high enough intensity of both transmitted and reflected light the metal film thickness has the order of the metal inverse absorption factor. The typical metal wavelike film thickness is about a few nanometers for the optical band. As an example, the aluminum half-permeable slab has thickness of 2 nm at 500 nm light wavelength. The experimental specimens of the OEs containing aluminum wavelike films were manufactured in collaboration with O.B. Serov using UV curing polymers. Both the holographic (locally periodic) type OEs and the OEs with chaotic diffuse film relief shape were produced. The process of OE fabrication was the following: firstly, the polymer replica of an object surface was produced. In one case, the surface of conventional reflection holograms with a phase relief was used as the object surface. In the other case, the chaotic diffuse surface was used. Then, a thin aluminum layer was sputtered onto the polymer replica. After that, a polymer slab was placed on top of the aluminum layer and covered by a piece of glass. We estimated the maximum phase modulation for the typical parameters of these samples: $h$=0.2 μm, $d_1$=0.002 μm, $n_0$=1.5, $n_1$=0.79, $\lambda$=0.5 μm, $\Lambda$=2 μm (for holographic wavelike film). Here we use the refractive index of aluminum from [27] for $n_1$. Substitution of these parameters into (27) gives $\delta$=0.007$\pi$. Thus, the condition (35) is carried out for the metal single-layer wavelike film with coefficient $b$ being less than 0.01. On the contrary, in conventional reflection holograms with phase relief the OPL of the rays passing through the hologram are varied up to $\Delta h(n_0-1)$, where $\Delta h$ is the change of depth of the surface relief, and $n_0$ is the index of refraction of the hologram. For this case typically $\Delta h = \lambda/4$ and $\delta \sim 0.25\pi$. This leads to significant phase modulation of the light passing through the holograms and consequently to light distortion.

The general expression (35) for maximum phase modulation is applicable for any type of the wavelike film relief shape within used limits. However, in case of a chaotic diffuse wavelike film relief some statistical characteristics of this relief should be taken into account to obtain a specific value of $\delta$. Such consideration exceeds the scope of this paper. Nevertheless, the principle of transmitted light distortion reduction is generally valid for the OE containing chaotic diffuse wavelike film. This was demonstrated by the following experimental study. We used the specimen containing some domains with and some domains without the upper polymer slab placed on top of a chaotic diffuse wavelike film. We measured the intensity of scattered light transmitted through that sample. The laser beam with 0.6 μm wavelength was directed onto the sample perpendicularly to its surface. The angular diagram of scattered light behind the sample is shown in Fig. 5. Here the relative intensity $I$ of scattered light in certain direction is plotted as a function of the angle $\nu$ between that direction and the normal to the sample surface. The relative intensity $I$ is determined as the scattered light intensity measured by photodetector divided by the incident beam intensity measured by the same detector. Curves A and B relate to the domains of the sample with and without the upper polymer slab respectively. It can be seen that the intensity of the directly transmitted beam ($\nu$ is equal 0) for curve A is almost 30-fold higher than one for curve B, the corresponding amounts are equal 0.83 and 0.029 respectively. Vice versa, the intensity of scattered light for curve A is substantially lower than the one for curve B for the majority of angles.

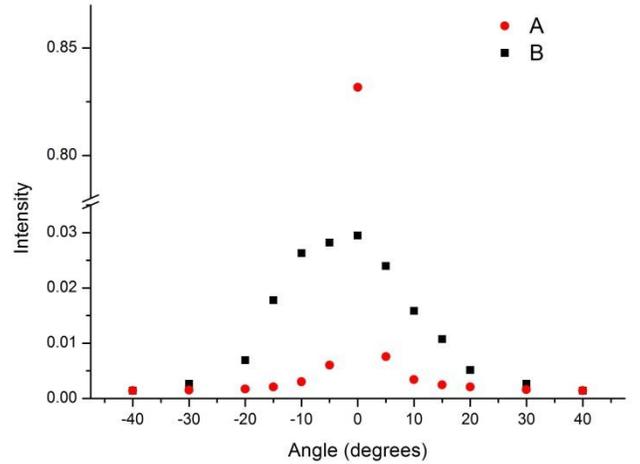

Fig. 5. Angular diagram of experimentally measured intensity of scattered light transmitted through the sample with the chaotic diffuse relief film. Relative intensity $I$ of scattered light in a particular direction is plotted as a function of the angle between this direction and the normal to the sample surface. Curve A relates to the OE with the transparent material placed below and above the wavelike film. Curve B relates to the OE with the transparent material placed only below the wavelike film.

Now let us consider the OE containing a multilayer film (Fig. 1 (c)). As follows from Snell's law the incident beam tilting angles inside every layer of the structure will not change if the structure would be replaced by the one with additional spaces with refraction indexes $n_0$ between adjacent layers. This allows generalizing the case of the multilayer wavelike film by adding the phase shifts of the beam in each layer. Then expression (35) becomes:

$$\delta = \frac{\pi n_0}{\lambda}[(|\phi^2 - \psi^2|)_{max} - (|\phi^2 - \psi^2|)_{min}]\sum_i \frac{d_i(n_i - n_0)}{n_i}, \qquad (39)$$

where $d_i$ and $n_i$ are the thickness and refractive index of $i$-th layer of the multilayer wavelike film respectively. For the case of sinusoidal shape of wavelike film (1) with two types of alternate layers with thicknesses $d_1$ and $d_2$ and refractive indexes $n_1$ and $n_2$ and normal incidence we can obtain from (38, 39):

$$\delta = \frac{4\pi^3 h^2 n_0 N}{\Lambda^2 \lambda}\left[\frac{d_1(n_1 - n_0)}{n_1} + \frac{d_2(n_2 - n_0)}{n_2}\right], \qquad (40)$$

where $N$ is the full number of double alternate layers. The terms within square brackets in (40) have the opposite signs and partly compensate each other. In case of multilayer film for optical wave band, different types of dielectrics could be used as alternate layers. The spacing $\Delta = d_1 + d_2$ of multilayer film in axis $z$ direction can be estimated from the Bragg condition $2d\sin\theta_B = \lambda/n_{av}$, where $n_{av}$ is an average refractive index of multilayer film, $\theta_B$ is a Bragg angle, (in our notations in Bragg case $\xi = \pi/2 - \theta_B$, see equation (9.10) in [28] or equation (14) in [25]). So, the thickness of each layer of a multilayer film significantly exceeds the thickness of the metal single-layer film. The entire film thickness will be at least one or two orders of magnitude higher than the one in a single-layer case. The phase modulation of transmitted light should be specially investigated in that case. This will be done in the next Section for a new type of color hologram, where equation (40) is used.

## 4. DISCUSSION: POSSIBLE APPLICATIONS

Let us now discuss the different types of OEs containing wavelike films and their possible applications. Several applications were mentioned in Introduction. These are head-up and head-mounted displays. For these applications, the holographic (locally periodic) wavelike films with optical power may be used. Bellow we will show that it is possible to make color displays by this technique.

The application of chaotic diffuse relief wavelike film is based on the fact of its diffuse scattering of only the reflected light. At the same time, the light transmitted through the diffuse wavelike film is not scattered and its phase is not distorted. The relief shape of the film can include both microasperities and macrocells, which are visible by a naked eye. The relief surface can be produced as a replica of the relief of the diffuse reflective object surface. The process of replica production here utilizes the source surfaces containing some relief drawing or pattern, such as engraved metal surfaces, fabrics, etc. The resultant element perfectly imitates the source surface texture.

The diffuse wavelike film can be used instead of a tinted car window. By choosing an appropriate thickness of the metal wavelike film, it is possible to get the element with any value of transmission factor. On the other hand, the selection of surface pattern may provide a required level of passenger compartment invisibility from outside for higher value of transmission factor as compared with conventional tinted windows in cars. Thus, the diffuse wavelike film is able to provide better traffic safety. In addition, this opens new prospects of car design. Similarly, the diffuse wavelike film may be used instead of mirror-glass windows in high-rise buildings. In this case the diffuse images forming by reflection may serve as elements of building design or can be used for outdoor advertising. Alternatively, the diffuse wavelike film can be used as window curtains. In this case, the fabric surface should be used as a source for replica production.

The multilayer diffuse wavelike film modeling the fabric surface texture may be used to create special effects. It is possible to manufacture an element that possesses spectral selectivity. Such an element may be visible only when illuminated with light of spectral band that intersects with the element spectral selectivity band. Otherwise, the element is invisible. This element may be employed in theatrical performances and shows to have a drop-curtain or actor's clothing details appear or disappear as a subject to illumination.

Another possible application of the diffuse wavelike film is image creation on spectacle lenses [9]. For example, it is possible to make a manufacturer's logotype visible to an outside observer but invisible for a spectacle user. In another case, the image formed by diffuse wavelike film can be an element of spectacle design. The method can be applied to both sunglass and vision correction lenses.

Multilayer wavelike films may also be used to make a high quality color hologram. We propose a color hologram representing a combination of three multilayer wavelike holograms recorded by different wavelengths (red, green, and blue) embedded into a single transparent medium (Fig. 6). The period in $z$ direction for each multilayer wavelike hologram is chosen to make it selectively reflective to the same wavelength, which was used for its recording. The color hologram is reconstructed by white light. Each of three multilayer wavelike holograms reconstructs the reflection image by light of the wavelength within the band of the multilayer wavelike film spectral selectivity. Reconstructing and reconstructed light at wavelengths outside of the hologram's spectral selectivity band propagates through the wavelike hologram without substantial phase distortion. That way the high quality color image is reconstructed. This hologram can be also used as a combiner of color head-up and head-mounted displays.

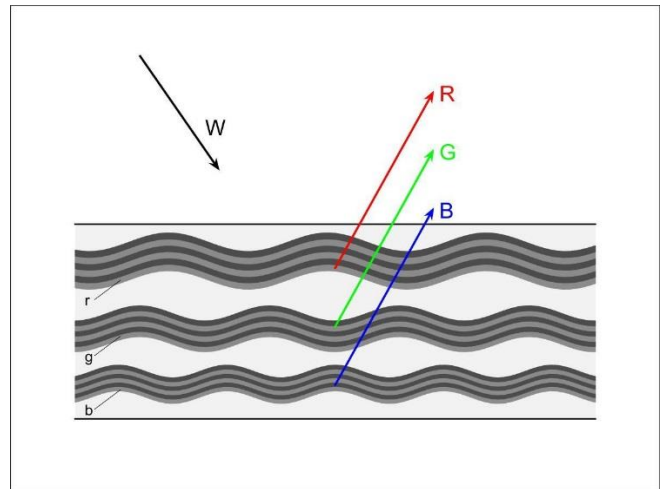

Fig. 6. Scheme of proposed color hologram. Three multilayer wavelike holograms r, g, and b recorded by different wavelengths (red, green, and blue respectively) are embedded into a single transparent medium. **W** is the reconstructing beam of white light; **R, G**, and **B** are reconstructed red, green, and blue light beams respectively.

The proposed color hologram has an advantage if compared with the known types of color holograms by the following reasons. One of the well-known color hologram types is the Denisyuk (also called Lippman–Bragg) hologram [29]. Such a hologram contains volume Bragg gratings with high spectral selectivity recorded in the bulk of the recording medium. The hologram reflects a narrow band around the recording wavelength from the reconstructing light spectrum. It is possible to obtain a high quality color image by recording of a superposition of three of such holograms in the same medium by light with three different wavelengths. However, the superposition of several holograms in the same medium decreases their diffraction efficiency (see section 17.6.3 in [28]). Therefore, the extra bright sources are required for the hologram reconstruction. Besides, each volume hologram can be produced only by the holographic recording process. Their mass-production by the embossing technology is impossible making such holograms to be expensive. Another color hologram type is the Benton (also called rainbow) hologram [30]. In Benton holography, the images with different wavelengths can be seen under different angles in a vertical plane. The correctly reconstructed image is seen only at a strict observer's eyes position. The Benton holograms have wide spread occurrence due to the possibility of their mass-production with low copy cost. However, the image quality of the Benton holograms is significantly lower than the one of the Denisyuk holograms, because of the use of the OEs (lenses and slits) during the master-hologram recording in the Benton's method. The two-steps hologram recording process (as a rule) is an additional disadvantage. Besides, the color of an image reconstructed from these holograms is changing with the change of observer's eyes position in a vertical plane. Thus, the observer sees the reconstructed image in relative colors.

The wavelike holograms may be mass-produced with the use of the known technology operations: recording of the master hologram on photoresist, photoresist developing, metallization, obtaining of the master stamp in the galvanoplastic process, and embossing on polymer layer [5, 6]. The next step should be deposition of the reflecting film onto the embossed relief. For single-layer wavelike film a thin (several nanometers) metal layer (aluminum or gold, for example) should be deposited. Alternatively, premetallized polymer may be used for embossing. To make a multilayer wavelike film a sequence of materials (preferably dielectrics) with different indexes of reflection should be deposited similarly to the technology of dielectric mirrors production. Then the same polymer material as the one, on which the relief picture was embossed should be put above. Its upper surface should be made plane. For color holograms all these steps should be repeated three times. Another way to manufacture the color hologram is to paste together three wavelike holograms (red, green, and blue). Both of these ways of color hologram manufacturing are similar to the corresponding technologies developed for multiple-data-layer optical disks production.

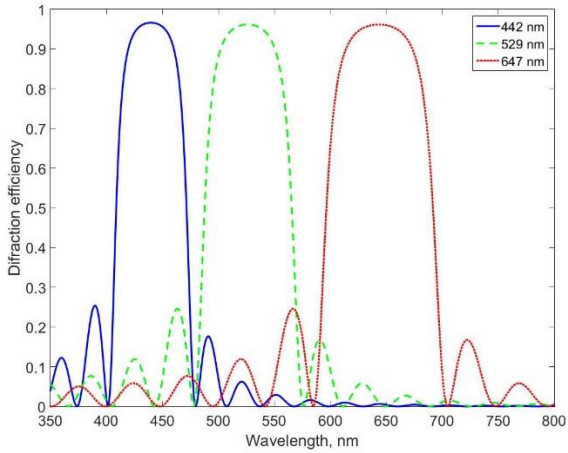

Fig. 7. Calculated spectral selectivity of color hologram containing three multilayer holographic wavelike films corresponding to the Bragg wavelengths 442 nm, 529 nm, and 647 nm (showed online by blue, green, and red curves respectively). The parameters of the hologram common for three multilayer holographic wavelike films are: $n_1 = 1.65$, $n_2 = 1.3812$, $N=13$, $c_R = -0.8102$, $c_1 = 0.6940$. The parameters of the hologram specific to each of three multilayer holographic wavelike films are: $K= 2\pi/\Lambda$ is equal to $2.8945\times10^6$ m$^{-1}$, $2.4184\times10^6$ m$^{-1}$, and $1.9774\times10^6$ m$^{-1}$, $d_1 = d_2$ are equal to $9.6561\times10^{-8}$ m, $1.1557\times10^{-7}$ m, and $1.4135\times10^{-7}$ m, $h$ is equal to $5.2\times10^{-8}$ m, $6.5\times10^{-8}$ m, and $7.5\times10^{-8}$ m for the multilayer holographic wavelike films corresponding to the Bragg wavelengths 442 nm, 529 nm, and 647 nm respectively.

The possibility of realization of the proposed type of color hologram was investigated by computer simulations. The structures containing three multilayer wavelike holograms corresponding to the Bragg wavelengths of 442 nm, 529 nm, and 647 nm with two types of alternate layers with refractive indexes $n_1$ and $n_2$ were studied. The calculations with different values of $n_1$ and $n_2$ were produced. We achieved good selective curves separation with $n_1 = 1.65$, $n_2 = 1.3812$ and $N=13$. This result is shown in Fig. 7. The other parameters were chosen based on conditions that the angles of incident and diffracted waves for each of three holograms are the same and have cosines $c_R = -0.8102$ and $c_1 = 0.6940$ respectively. These angles define the wave number $K= 2\pi/\Lambda$ of each wavelike hologram in $x$ direction. For the holograms corresponding to the wavelengths 442 nm, 529 nm, and 647 nm $K$ are equal to $2.8945\times10^6$ m$^{-1}$, $2.4184\times10^6$ m$^{-1}$, and $1.9774\times10^6$ m$^{-1}$ respectively. The periods of the holograms in $z$ direction were found from the Bragg condition. We assumed that $d_1 = d_2$. Then $d_1$ is equal to $9.6561\times10^{-8}$ m, $1.1557\times10^{-7}$ m, and $1.4135\times10^{-7}$ m for the wavelike holograms corresponding to wavelengths 442 nm, 529 nm, and 647 nm respectively. In this calculation, we used values of $h$ equal to $5.66\times10^{-8}$ m, $6.77\times10^{-8}$ m, and $8.28\times10^{-8}$ m for the wavelike holograms corresponding to wavelengths 442 nm, 529 nm, and 647 nm respectively. We assume the refractive index $n_0$ of the plane-parallel medium containing wavelike holograms to be equal to the average hologram refractive index. For the values used in calculation shown in Fig. 7 $n_0 =1.5215$. Now let us estimate the maximum phase modulation of light transmitted through the structure, which we are studying. After substitution of the parameters values listed above into expression (40) we obtain $\delta$ equal to $-0.0027\pi$ for each of three holograms. This value of the maximum phase modulation of transmitted light is small enough to consider the level of transmitted light distortion to be negligible. Thus, we have proven a principal realization possibility of the proposed color hologram.

In summary, the dynamic theory of light diffraction by a multilayer locally periodic semitransparent wavelike film was developed. A simple analytical formula for near Bragg diffraction order intensity was obtained for a case when only one diffraction order lies within the hologram angular selectivity. The phase modulation of light transmitted through an OE containing semitransparent wavelike films was estimated for single layer and multilayer wavelike films with an arbitrary shape of surface. These results were used to calculate distortion of light transmitted through a holographic wavelike film. The scattering of light transmitted through the OE containing wavelike film having chaotic diffuse relief shape with transparent material lower and above/only lower wavelike film was measured. A new type of high quality color holograms was proposed. Its feasibility was proven by calculation of hologram spectral selectivity and by estimation of transmitted light distortion for three colors.

**Acknowledgment**. This work was done in the network of the Memorandum of understanding between Kotel'nikov Institute of Radio Engineering and Electronics of the RAS, Moscow, Russia and University of Sonora in Hermosillo, Sonora, Mexico.

The experimental specimens of the OE containing semitransparent wavelike films were obtained some time ago together with O.B. Serov (deceased in 2004). We thank S.G. Zybtsev for providing laser equipment used to measure scattering by the OE. We thank D.A. Oulianov and S. Lobassov for their help with text editing.